\documentclass[12pt]{JHEP3}
\usepackage{graphicx}%
\usepackage{epsfig}
\epsfclipon
\usepackage{multicol}
\usepackage{epsfig,bm}
\usepackage{amssymb,amsmath}
\newcommand{\roughly}[1]{\mathrel{\raise.3ex\hbox{$#1$\kern-0.85em
\lower1ex\hbox{$\sim$}}}}

\newcommand{\bmat}{\left(\begin{array}}
\newcommand{\emat}{\end{array}\right)}

\def\yzero{\smash{\hbox{$y\kern-4pt\raise1pt\hbox{${}^\circ$}$}}}

\def\beq{\begin{equation}}
\def\eeq{\end{equation}}
\def\beqa{\begin{eqnarray}}
\def\eeqa{\end{eqnarray}}

\def\-{\hphantom{-}}

\def\s2{\frac{1}{2}}

\def\IF{\relax{\rm I\kern-.18em F}}
\def\II{\relax{\rm I\kern-.18em I}}
\def\IP{\relax{\rm I\kern-.18em P}}
\def\IC{\relax{\rm I\kern-.48em C}}
\def\IR{\relax{\rm I\kern-.18em R}}

\def\cG{{\cal G}}

\def\cP{{\cal P}}
\def\cW{{\cal W}}

\def\Dsl{\,\raise.15ex\hbox{/}\mkern-13.5mu D} 
\def\IZ{{\Bbb Z}}
\def \one{\relax{\rm 1\kern-.26em I}}

\def\bt{\overline{\tau}}
\def\gvw{{\scriptscriptstyle GVW}}
\def\exd{{\rm d}}
\def\nn{\nonumber}

\def\pref#1{(\ref{#1})}

 \def\cp#1{\relax\ifmmode {\IP\kern-2pt{}_{#1}}\else $\IP\kern-2pt{}_{#1}$\=fi}

\catcode`\@=11
\newdimen\@rotdimen
\newbox\@rotbox

\def\@vspec#1{\special{ps:#1}}
\def\@rotstart#1{\@vspec{gsave currentpoint currentpoint translate
   #1 neg exch neg exch translate}}
\def\@rotfinish{\@vspec{currentpoint grestore moveto}}
\def\@rotr#1{\@rotdimen=\ht#1\advance\@rotdimen by\dp#1%
   \hbox to\@rotdimen{\hskip\ht#1\vbox to\wd#1{\@rotstart{90 rotate}%
   \box#1\vss}\hss}\@rotfinish}
\def\@rotl#1{\@rotdimen=\ht#1\advance\@rotdimen by\dp#1%
   \hbox to\@rotdimen{\vbox to\wd#1{\vskip\wd#1\@rotstart{270 rotate}%
   \box#1\vss}\hss}\@rotfinish}%
\def\@rotu#1{\@rotdimen=\ht#1\advance\@rotdimen by\dp#1%
   \hbox to\wd#1{\hskip\wd#1\vbox to\@rotdimen{\vskip\@rotdimen
   \@rotstart{-1 dup scale}\box#1\vss}\hss}\@rotfinish}%
\def\@rotf#1{\hbox to\wd#1{\hskip\wd#1\@rotstart{-1 1 scale}%
   \box#1\hss}\@rotfinish}%
\def\rotate{\@ifnextchar[{\@rotate}{\@rotate[l]}}
\def\@rotate[#1]#2{\setbox\@rotbox=\hbox{#2}\@nameuse{@rot#1}\@rotbox}

\catcode`\@=12

\title{Nonrenormalization of Flux Superpotentials in String
Theory}

\author{C.P. Burgess,${}^{1,2}$ C. Escoda${}^3$
and F. Quevedo${}^3$\\

${}^1$ Department of Physics and Astronomy, McMaster University,\\
\qquad 1280 Main Street West, Hamilton, Ontario, Canada, L8S 4M1.\\
${}^2$ Perimeter Institute, 31 Caroline Street North,\\ \qquad
Waterloo, Ontario, Canada.\\
${}^3$ Centre for Mathematical Sciences, DAMTP, Cambridge
University,\\ \qquad Cambridge CB3 0WA UK.

}

\abstract{Recent progress in understanding modulus stabilization
in string theory relies on the existence of a non-renormalization
theorem for the 4D compactifications of Type IIB supergravity
which preserve $N=1$ supersymmetry. We provide a simple proof of
this non-renormalization theorem for a broad class of Type IIB
vacua using the known symmetries of these compactifications,
thereby putting them on a similar footing as the better-known
non-renormalization theorems of heterotic vacua without fluxes.
The explicit dependence of the tree-level flux superpotential on
the dilaton field  makes the proof more subtle than in the absence
of fluxes.}

\preprint{DAMTP-2005-82}

\begin{document}

\makeatletter \@addtoreset{equation}{section} \makeatother
\renewcommand{\theequation}{\thesection.\arabic{equation}}

\setcounter{page}{1} \pagestyle{plain}
\renewcommand{\thefootnote}{\arabic{footnote}}
\setcounter{footnote}{0}


\section{Introduction}

Four-dimensional theories with $N=1$ supersymmetry are completely
characterized at low energies by three functions --- a K\"ahler
potential $K(\varphi,\varphi^*)$, a superpotential $W(\varphi)$
and a gauge-kinetic function $f_{ab}(\varphi)$. The arguments of
these functions are the complex scalar fields, $\varphi^i$, which
appear within the chiral matter multiplets of these theories, and
supersymmetry dictates that both $W$ and $f_{ab}$ are holomorphic
functions of these arguments.

Although the K\"ahler function receives corrections order-by-order
in perturbation theory in these theories, the superpotential does
not and the gauge-kinetic function typically only does at one
loop. These properties for $W$ and $f_{ab}$ are known as
non-renormalization theorems, and because $W$ controls the vacua
of the theory they play a crucial role in understanding these
vacua and in particular the circumstances under which
supersymmetry can spontaneously break.

These theorems were originally proven using the detailed
properties of supersymmetric perturbation theory about flat space
\cite{NRTheorems}, but a more robust understanding was achieved
with the derivation of the non-renormalization results as a
consequence of symmetry arguments combined with the holomorphic
dependence on $\varphi^i$ which supersymmetry requires of $W$ and
$f_{ab}$ \cite{ds,BFM,seiberg,NRTheoremsSym}. Among the advantages of the
symmetry formulation is the ability easily to extend the results
to curved spacetimes and to string theory.

\subsection{Non-renormalization for String Vacua}

For four-dimensional compactifications of heterotic vacua which
are $N=1$ supersymmetric the non-renormalization argument for $W$
is very simple to state \cite{ds}. It is based on the observation
that for these vacua the two string perturbative expansions --- in
the string coupling, $\lambda_s$, and in low-energies, $\alpha'$
--- correspond in the supersymmetric low-energy theory to an
expansion in powers of two of the effective theory's scalar
fields. These fields are the 4D dilaton, $e^\phi$, and the field
$\sigma$ which describes the volume of the underlying Calabi-Yau
compactification. For heterotic vacua these fields combine with
two low-energy axions into two complex fields, $S$ and $T$, which
are the scalar components of two matter supermultiplets, and at
lowest order a direct dimensional reduction shows that neither of
these fields appears in the low energy superpotential
\cite{truncation,BFM}.

The non-renormalization argument then proceeds as follows. At
higher orders the dependence of $W$ and $f_{ab}$ on $S$ and $T$ is
dictated by holomorphy and low-energy symmetries \cite{BFM,ds}. In
particular, there is a low-energy Peccei-Quinn (PQ) symmetry $S
\rightarrow S + i\omega$ (where $\omega$ is a constant parameter),
which forbids $W$ from depending on $\hbox{Im}\,S$ to all orders
in perturbation theory. However since $W$ must be a holomorphic
function of $S$, this also precludes $W$ from depending on
$\hbox{Re}\,S \propto e^\phi$, which is the string coupling
constant. We conclude from this that the PQ symmetry precludes $W$
from developing a dependence on $S$, and so ensures that $W$
receives no corrections in string perturbation theory.

The $\alpha'$ expansion similarly involves the field $\sigma$
contained within $\hbox{Re}\,T$, and a similar argument involving
another axion-like symmetry which shifts $\hbox{Im}\,T$ also
precludes $W$ from acquiring a dependence on $T$. The same
symmetry allows $f_{ab}$ to receive $T$-dependent contributions,
but only at one string loop. It precludes corrections to $f_{ab}$
from arising at higher loops. Arguments such as these show how
symmetries can imply the non-renormalization of $W$ and allow only
a limited renormalization of $f_{ab}$. Their implications are
typically restricted to perturbation theory because
nonperturbative effects can break the underlying symmetries. Even
in this case consistency with the known anomalies in the
underlying axionic symmetries requires that the dependence
acquired by $W$ must be purely exponential in $S$ and $T$.

In recent years Type IIB vacua have become the focus of much
attention, due to their utility in addressing the long-standing
problem of the stabilization of string moduli \cite{KKLT}. $N=1$
supersymmetric 4D compactifications of Type IIB vacua are useful
for modulus stabilization because in the presence of background
fluxes they generate a superpotential, $W$, which can stabilize
all of the complex-structure moduli of the underlying warped
Calabi-Yau geometry \cite{GKP} (see also \cite{sethi}). For
compactifications with constant dilaton this superpotential has a
specific (Gukov-Vafa-Witten) form \cite{gvw}, $W_{\gvw}$, which
allows a simple expression (see eq.~\pref{GVWForm}) in terms of
geometrical quantities defined on the Calabi-Yau internal space.
For $F$-theory compactifications for which the dilaton varies
across the internal dimensions the superpotential is more
complicated, but is also known \cite{gvw, GVWFTheory}.

What is important about the Type-IIB moduli-stabilization
arguments of ref.~\cite{KKLT} is that they are done in a
controllable way, within the domain of validity of the $\lambda_s$
and $\alpha'$ expansions. The non-renormalization of $W$ plays an
important part in doing so, since it is what precludes the
appearance of corrections to $W$ order-by-order in $\lambda_s$ and
$\alpha'$. Unfortunately, a justification of the
non-renormalization theorem for Type-IIB vacua along the lines
used for heterotic vacua has not yet been made, largely because
symmetries cannot preclude $W_\gvw$ from depending on the string
coupling due to the explicit presence on the dilaton, $e^\phi$,
which is already present in the lowest-order superpotential,
$W_\gvw$. We note for later purposes that although this dilaton
dependence complicates the non-renormalization argument for the
string-coupling expansion, we {\it do} know that $W_\gvw$ does not
receive any corrections within the $\alpha'$ expansion just as for
heterotic vacua, because it cannot depend explicitly on the
K\"ahler-structure moduli.

Our purpose in this note is to fill in this step by providing, in
\S2 below, a simple derivation of the non-renormalization theorem
for Type IIB vacua based only on holomorphy and symmetries.
Besides filling in a missing step in the modulus-stabilization
arguments this renormalization theorem can also have other
applications, some of which we briefly describe in \S3.

\section{Non-Renormalization for Type IIB Vacua}

In this section we explain how the non-renormalization theorem
follows from holomorphy and other incidental symmetries of the
low-energy effective 4D action. We do so by starting with a brief
summary of the low-energy field content and a description of the
low-energy symmetries on which our arguments are based.

\subsection{10D Type IIB Supergravity}

The massless sector of Type IIB string compactifications is
described by ten-dimensional Type IIB supergravity, which has two
supersymmetries generated by two Majorana-Weyl supercharges which
share the same 10D chirality. This supersymmetry algebra also
enjoys a $U(1)$ $R$-symmetry which rotates these two supercharges
into one another. The field content of the theory consists of the
graviton, $g_{MN}$, with $R$-charge $q=0$; a complex Weyl
gravitino, $\psi_M$, with $q = 1/2$; the NS-NS and R-R 2-forms,
$B^1_{MN}$ and $B^2_{MN}$, with $q=1$; a complex Weyl dilatino,
$\lambda$, with $q = 3/2$; two scalars (the dilaton, $e^\phi$, and
the R-R 0-form, $C_{(0)}$), with $q=2$; and the self-dual 4-form
$C_{(4)}$, with $q=0$.

The bosonic part of the 10D low-energy effective theory for the
Type IIB sector has the following form in the string frame
\cite{GKP}, to lowest order in $\alpha'$ and string coupling:
\beqa \label{BoseSugra}
    S_{\rm S} &=& \frac12 \int d^{10}x \; \sqrt{-g_s} \left\{ e^{-2\phi}
    \Bigl[  R_s + 4 (\partial \phi)^2 \Bigr] - \frac12 \, F_{(1)}^2
    - \frac{1}{12} \, G_{(3)} \overline{G}_{(3)} - \frac{1}{480} \,
    {F}_{(5)}^2 \right\} \\
    && \qquad -\frac{i}{8} \int e^\phi \, C_{(4)} \wedge G_{(3)}
    \wedge \overline{G}_{(3)} + \sum_b \left[- \mu_b \int_b
    d^{(p+1)}y
    \, e^{-\phi} \sqrt{-\hat{g_s}} + \mu_b \int_b  \hat{\cal C}_{(p+1)}
    \right] \,, \nonumber
\eeqa
where $F_{(p+1)} = \exd C_{(p)} +\cdots$, $\tau = C_0 + i e^{-
\phi}$, $H_{(3)}^\alpha = \exd B^\alpha_{(2)}$ and $G_{(3)} = \tau
\, H^1_{(3)} + H^2_{(3)}$. The sum on `$b$' runs over all of the
various D$p$-branes which source the bulk fields, and $\hat{g}$
and $\hat{\cal C}$ denote the pull-back to the brane world-sheet
of the bulk metric and the appropriate $(p+1)$-form which defines
the brane Wess-Zumino interaction with the RR potentials.

It is useful to reformulate the above action \pref{BoseSugra} by defining
the Einstein metric $g_{MN} = e^{-\phi/2} g_{sMN}$, and therefore the action
becomes
\beqa \label{AIIB}
S_{\rm E} &=& \frac12 \int d^{10}x \; \sqrt{-g} \Biggl\{ R -
\frac{\partial_M\tau \partial^M \tau}{2 ({\rm Im}\, \tau)^2} -
 \frac{{G_{(3)} \cdot \bar{G}_{(3)}}}{
12\, {\rm Im}\, \tau } - \frac{1}{480} \,  {F}_{(5)}^2 \Biggr\} \\
&& -  \frac{i}{8} \int  \frac{C_{(4)} \wedge G_{(3)} \wedge \bar{G}_{(3)}
}{ {\rm Im}\, \tau}\ +\ \sum_b \left[- \mu_b \int_b
    d^{(p+1)}y
    \, e^{(p-3) \phi/4} \sqrt{-\hat{g}} + \mu_b \int_b  \hat{\cal C}_{(p+1)}
    \right] . \nonumber
\eeqa
At the classical level the bulk part of this theory enjoys an
accidental global $SL(2,R)$ invariance which plays an important
role in our arguments. This symmetry is nonlinearly realized, with
the scalars $\phi$ and $C_0$ transforming as a coset
$SL(2,R)/U(1)$\cite{sch,sw,berh,bcg,dalla,keh, michael}. In terms
of the complex combination, $\tau$, defined above, the action of
this symmetry takes the standard form
\beqa \label{tauTransformation}
    &&\tau \rightarrow \frac{a\tau+b}{c\tau+d}\, , \   \  \ \ \ \
    \ B^\alpha_{MN}\rightarrow
    {(\Lambda^T)^{-1}}^\alpha_{\mbox{\phantom{a}}\beta}
    B^\beta_{MN} \,\ \ \ \ \ \Lambda=
    \left(\begin{array}{cc}a & b \\c & d\end{array}\right)
    \!\in\! SL(2,{\bf R})\, , \nonumber \\
    && \qquad \qquad \qquad
    g_{MN}\rightarrow g_{MN}\, , \   \  \  \  \  \  \  \  \
    C_{MNPQ}\rightarrow C_{MNPQ}\, , \nonumber \\
\eeqa
while the fermions transform as
\beq
    \psi_M \rightarrow \left( \frac{c \bt+d}{c \tau+d}
    \right)^{1/4} \psi_M \, \  \  \  \  \ \lambda \rightarrow
    \left(\frac{c \bt+d}{c\tau+d}\right)^{3/4}\lambda \,.
\eeq
As usual the real parameters $a$ through $d$ satisfy $ad-bc = 1$.

The field strengths for the bosonic fields can be combined into
combinations which also transform simply under the global
$SL(2,R)$ transformations, as follows. In terms of $H_{(3)}^\alpha
= \exd B^\alpha_{(2)} $ and $F_{(5)} = \exd C_{(4)} + \dots$, the
combinations
\beq \label{PQGDefs}
    P_M = i \, \frac{\partial_M \tau}{2\, \tau_2} \,,\quad
    Q_M = - \frac{\partial_M \tau_1}{2 \, \tau_2} \qquad
    \hbox{and} \qquad
    G_{(3)} =
    \tau \, H^1_{(3)}
    + H^2_{(3)}
\eeq
inherit the following transformation properties under $SL(2,R)$:
\beq \label{G3Transformation}
    P_M \to \left( \frac{c \bt + d}{c \tau + d} \right) \, P_M \,,
    \quad
    Q_M \to Q_M - \frac{i}{2} \, \partial_M \ln \left( \frac{c \bt
    + d}{c \tau + d} \right) \,, \quad
    G_{(3)} \to
    \, \frac{G_{(3)}}{c\tau + d}
\eeq
and $F_{(5)} \to F_{(5)}$.

The $R$ symmetry can be regarded as a subgroup of $SL(2,R)$, as
may be seen by choosing the following one-parameter family of
$SL(2,R)$ transformations:
\beq \label{SL2R}
    b = -|\tau| \,, \quad
    c = 1/|\tau| \quad \hbox{and} \quad
    d = 0 \,,
\eeq
for any given $\tau$. These choices satisfy $ad - bc = 1$ and have
the property that $c \tau + d = \tau/|\tau| = e^{-i \alpha}$,
where $\alpha = - \hbox{arg} \, \tau$. With this choice $c \bt + d
= e^{+i\alpha}$ and so the $SL(2,R)$ transformation properties of
the field strengths and fermions imply these transform under this
one-parameter subgroup as
\beqa
    &&P_M \to e^{+2i\alpha} \, P_M \,, \quad
    Q_M \to Q_M + \partial_M \alpha \,,\quad
    G_{(3)} \to e^{+i\alpha} \, G_{(3)} \,, \nn\\
    &&\psi_M \to e^{+i\alpha/2} \, \psi_M \,
    \quad \hbox{and} \quad
    \lambda \rightarrow
    e^{+3i\alpha/2} \, \lambda \,,
\eeqa
which may be recognized as the $R$ transformation properties for
each of these fields.

The normalization of the $R$ charge may be obtained by seeing how
the supersymmetry parameter, $\epsilon$, transforms, which may be
inferred from the above expressions together with the
supersymmetry transformation rules
\cite{sch,sw,berh,bcg,dalla,keh}:
\beqa
    \delta \lambda &=& \frac{i}{\kappa} \Gamma^{M} P_{M}
    \epsilon^{*} - \frac{1}{24\sqrt{\tau_2}} \,
    \Gamma^{MNP} G_{MNP} \,\epsilon
    \nn\\
    \delta \psi_M &=&  \frac{1}{\kappa} D_M  \epsilon + \frac{i}{480}
    \Gamma^{M_1 ... M_5} F_{M_1 ... M_5} \Gamma_M \,\epsilon \\
    &&\qquad \qquad -
    \frac{i}{96 \sqrt{\tau_2}} \left( \Gamma_{M}^{PQR} G_{PQR}
    - 9 \Gamma^{PQ}
    G_{MPQ} \right) \epsilon^{*} \,, \nn
\eeqa
where
\beq
    D_M \epsilon = \left( \partial_M + \frac{1}{4} \omega_M^{AB}
    \Gamma_{AB} - \frac{i}{2}Q_M \right) \epsilon \,.
\eeq
We read off from these expressions that $\epsilon$ has $R$ charge
$q = \frac12$.

\subsection{Symmetry Breaking}

Since our interest is in using these symmetries to constrain the
properties of the quantum-corrected superpotential and gauge
potential, it behooves us to understand which symmetries survive
perturbative quantum corrections. In particular, it is known that
the $SL(2,R)$ symmetry is only a symmetry to leading order in
$\lambda_s$ and $\alpha'$, which does not survive intact into the
quantum theory. For this reason we identify several important
subgroups of $SL(2,R)$ which do survive quantum corrections. The
three important such symmetries which we use in the following are:
\begin{enumerate}
\item {\bf R-Invariance:} We have seen that the $U(1)$ $R$
symmetry of the 10D supersymmetry transformations is a subgroup of
$SL(2,R)$, and is a symmetry at leading order in $\alpha'$ because
it is a symmetry of the 10D supergravity given by
eq.~\pref{BoseSugra}. This is not an exact symmetry of the string
theory, however, and there is evidence that this symmetry is
broken at subleading but finite order in the $\alpha'$ expansion
\cite{narain}. However if we restrict ourselves to leading order
in $\alpha'$, we now argue that this symmetry remains unbroken to
all orders in the string coupling, $e^\phi$. This follows because
10D supersymmetry completely dictates the dilaton dependence of
the two-derivative action, eq.~\pref{BoseSugra}, which gives the
action to leading order in $\alpha'$, showing that the
two-derivative terms of the 10D action are not renormalized in
string perturbation theory.
\item {\bf Peccei-Quinn (PQ) Invariance:} Another subgroup of
$SL(2,R)$ which is anomalous but which also survives to all orders
in string perturbation theory is the PQ symmetry defined by the
$SL(2,R)$ transformations $a = d = 1$ and $c = 0$, for which $\tau
\to \tau + b$.
\item {\bf $SL(2,Z)$ Invariance:} Although quantum corrections
generically break the $SL(2,R)$ invariance, they are expected to
preserve the discrete subgroup which is obtained if the parameters
$a$, $b$, $c$ and $d$ are restricted to be integers. This symmetry
is expected to survive but only after non-perturbative corrections
are included. It will not play an important role in our arguments
which are only perturbative.
\end{enumerate}

All three of these symmetries may be used to constrain the form of
the low-energy effective action to all orders in perturbation
theory. We now turn to a description of their implications for
compactifications whose low-energy action is described by an $N=1$
4D supergravity.

\subsection{The Effective 4D Supergravity}

It is always possible to use the standard supergravity action to
describe the dynamics of the low-energy degrees of freedom in any
compactification to four dimensions which preserve $N=1$
supersymmetry in 4 dimensions. The same is true for
compactifications which break 4D $N=1$ supersymmetry, provided the
4D particle supermultiplets of interest are split by less than the
Kaluza-Klein (KK) scale.\footnote{In general, an effective theory
can be described by the standard 4D supergravity provided that the
low-energy field content can linearly realize the approximate
$N=1$ supersymmetry. This is generically possible if the mass
splittings amongst the low-energy multiplets is smaller than the
masses of the heavy particles whose removal generates the
effective theory in question.}

There is a broad class of 4D compactifications of Type IIB
supergravity having an approximate $N=1$ supersymmetry which are
described by a low-energy 4D supergravity \cite{GKP}. The
low-energy fields for these compactifications consist of 4D
supergravity coupled to a number of $N=1$ gauge and matter
supermultiplets. The scalar fields of the matter multiplets
include the various compactification moduli, various low-energy
axion fields, and scalars of other types. Among these massless
fields is the volume modulus, $\sigma$, whose scalar component
contains the extra-dimensional volume field, $V$, as well as the
scalar which is related to $V$ by supersymmetry, obtained in four
dimensions by dualizing one of the components, $C_{\mu\nu pq}$, of
the 4-form field, $C_{(4)}$ which have two indices in the
uncompactified directions, $x^\mu$, $\mu = 0,..,3$.

As mentioned in the introduction, any 4D $N=1$ supergravity is
characterized by its K\"ahler function, $K$, superpotential, $W$,
and gauge kinetic term, $f_{ab}$. For non-renormalization theorems
our interest in particular is in the holomorphic functions $W$ and
$f_{ab}$. For Type IIB theories for which the dilaton field $\tau$
varies trivially over the extra dimensions the expression for the
superpotential to leading order in string and $\alpha'$
perturbation theory is given in terms of the 10D fields by the
Gukov-Vafa-Witten expression \cite{gvw}:\footnote{A generalization
of this expression applies to $F$-theory compactifications for
which $\tau$ varies across the extra dimensions.}
\beq \label{GVWForm}
    W_{\gvw} = \int_M G_{(3)} \wedge \Omega \,,
\eeq
where $G_3$ is as above and $\Omega$ is the unique $(3,0)$ form of
the Calabi-Yau space which underlies the internal six-dimensional
space, $M_6$. This defines a function of the 4D fields which
appear in the low-energy theory because $\Omega$ depends
implicitly on those 4D fields which correspond to the
complex-structure moduli of $M_6$ (but not on all such fields,
such as those associated with the K\"ahler moduli). Similarly,
$W_{\gvw}$ also depends on $\tau$ through the definition,
eq.~\pref{PQGDefs}, of $G_{(3)}$. It is the dependence on these
fields of the resulting scalar potential, $V_{\gvw}$, which
stabilizes the moduli --- including in particular $\tau$ ---
described by these fields.

It is instructive to exhibit explicitly the $SL(2,R)$ transformation
of this leading-order superpotential, which transforms in the same
way as does $G_{(3)}$:
\beq \label{WSL2RTransformation}
    W_{\gvw} \to \frac{W_{\gvw}}{c\tau + d} \,.
\eeq
This is as required to make the low-energy action invariant since
the variation of $W$ cancels the variation of the K\"ahler
function, $K$, which has the form
\beq
    K = - \ln(\tau - \bt) + \hat{K} \,,
\eeq
where $\hat{K}$ is an $SL(2,R)$ invariant function of the other
fields. Since the low-energy action only depends on $K$ and $W$
through the combination $K + \ln|W|^2$, the transformation of the
first term in $K$ precisely cancels the transformation of $W$,
given that
\beq
    \tau - \bt \to \frac{\tau - \bt}{|c\tau + d|^2} \,.
\eeq

At this point we can see why the arguments for the
non-renormalization of $W$ for heterotic vacua do not
straightforwardly apply to Type IIB vacua. Although there is a
shift symmetry, $\tau \to \tau + b$, this cannot preclude $W$ from
depending on $\tau$ because $W_{\gvw}$ already has a $\tau$
dependence through the $\tau$'s which enter into the definition of
$G_{(3)}$. This is possible because of the presence of the
background fluxes, $\langle H^1_{mnp} \rangle$ and $\langle
H^2_{mnp} \rangle$, which also transform under the shift
symmetry.\footnote{Notice that for topological (and therefore
non-perturbative) reasons, the fluxes are quantised breaking  the
$SL(2,R)$ symmetry to $SL(2,\IZ)$. Since our arguments are only
perturbative we will not make use of this constraint in our coming
discussion. We thank S. de Alwis and J. Conlon for helpful
discussions on this point.}

\subsection{The Non-Renormalization Theorem}

We now show how to use the symmetries given above to derive a
non-renormalization result for the Type IIB case. In order to do
so we assume that the low-energy 4D effective theory linearly
realizes $N=1$ supersymmetry, and so may be written in terms of
the standard $N=1$ supergravity action, characterized by the
functions $K$, $W$ and $f_{ab}$.

To start we first require a statement of the field content of the
effective 4D theory, and how these fields transform under the
global symmetries of the 10D theory. For these purposes we keep
track of two kinds of fields. The first kind consists of the
fields which describe the light degrees of freedom whose masses
are smaller than the KK scale, and so whose dynamics is described
by the low-energy $N=1$ supersymmetric 4D theory. The scalar
fields of this theory --- which we denote collectively as
$\varphi^i$ --- transform under supersymmetry as chiral matter. We
also imagine there to be a low-energy gauge group, with gauge
multiplets denoted by $\cW^a$, as well as the 4D supergravity
multiplet itself.

Included among the matter multiplets are both the
K\"ahler-structure moduli (including the volume modulus, $\sigma$)
which are left massless by $W_{\gvw}$, and the complex-structure
moduli (including the dilaton multiplet $\tau$) which appear
explicitly in $W_{\gvw}$. The complex-structure moduli can appear
in the low-energy effective theory because the masses they acquire
because of their presence in the GVW potential are systematically
light (for large extra-dimensional volume) compared with the KK
scale. These fields can be defined without loss of generality to
be invariant under the PQ symmetry, by absorbing into their
definition the appropriate power of $e^{i\tau}$.

The second class of fields whose dependence we follow in the
low-energy action are `spurions' \cite{Spurion}, which describe
the transformation properties of the background flux
vacuum-expectation-values ({\it v.e.v.}'s) under the symmetry
transformations of interest. These fluxes reside within the
background value for the 2-form, $G_{(3)}$, and may be regarded as
the {\it v.e.v.}s, $\cG^r$, of the large collection of 4D scalar
fields which are obtained when $G_{(3)}$ is dimensionally reduced.
As we have seen, these fields transform nontrivially under the $R$
and PQ symmetries, with $\cG^r$ transforming as in
eq.~\pref{G3Transformation}.

Both types of fields are important for the non-renormalization
theorem, because string perturbation theory is related to the
dependence of the low-energy action on $e^\phi$ (and so on
$\tau$), while the $\alpha'$ expansion is related to the
dependence on the volume modulus $\sigma$. It is also useful to
follow the dependence on the spurions $\cG^r$, since these contain
$\tau$ and transform under the $R$ symmetry.

To establish the non-renormalization theorem, we now argue in two
steps. First we restrict our attention to lowest-order in the
$\alpha'$ expansion, and argue that the superpotential is not
renormalized to all orders in string perturbation theory. For this
part of the argument the $R$ symmetry can be regarded to be
unbroken and so can be used to constrain the possible form for
$W$. We then separately argue that this result also remains true
to all orders in the $\alpha'$.

To leading order in $\alpha'$, but to all orders in string loops,
we may use the $R$-invariance of $W$ to restrict its form. We may
do so because although the $R$ symmetry is broken by string loops,
we have argued above that this breaking cannot arise to leading
order in $\alpha'$, due to the restrictive form of those terms in
the 10D supergravity action involving two or fewer derivatives.
One might worry that the $R$ symmetry might be spontaneously
broken by the background, if background fields (like fluxes) break
the $R$ symmetry. But this is the point of keeping the spurion
fields, $\cG^r$, which capture how the background fields
transform. Provided the fields $\cG^r$ are the only backgrounds
which break the $R$ symmetry, keeping track of their appearance in
the low-energy 4D theory guarantees this low-energy action is
$R$-invariant (to lowest order in $\alpha'$). The only bosonic
background field which can break $R$ invariance and yet is {\it
not} encoded in the $\cG^r$'s would be a background value for
$P_m$, and so our analysis does not cover fields for which the
dilaton field $\tau$ varies across the internal Calabi-Yau space.
By tracking only the dependence on $\cG^r$ we assume $P_m=0$, and
so restrict our analysis to those orientifold limits of $F$-theory
for which the Gukov-Vafa-Witten flux superpotential, $W_{GVW}$,
applies.

Because the supersymmetry transformation parameter has $R$-charge
$q_\epsilon = +\frac12$, it follows that $R$-invariance of the
action implies the superpotential must carry $R$-charge $q_W =
+1$. Since $G_{(3)}$ also carries $q_G = +1$ we are always free to
take $W$ to be proportional to one of the $\cG^r$s --- say $\cG^0$
--- and so write
\beq \label{WSymForm}
    W(\varphi^i,\tau; \cG^r) = \cG^0 \, w \left(
    \varphi^i; \frac{\cG^r}{\cG^0}\right) \,,
\eeq
for some function $w$. $w$ cannot depend separately on $\tau$
(beyond the $\tau$ dependence of the $\cG^r$'s) because $w$ must
be PQ-invariant and $\tau$ shifts under this symmetry while all of
the other arguments of $w$ do not transform. Explicit calculation
shows that the lowest-order GVW result, $W_{\gvw}$, corresponds to
$w$ being given by a strictly linear function of the arguments
$\cG^r/\cG^0$,
\beq \label{LOW}
    W_{\gvw}(\varphi^i, \tau;\cG^r) = \sum_{r\ge0} \cG^r
    \, w_r(\varphi^i) \,.
\eeq

To establish the non-renormalization theorem for the
string-coupling expansion, we now argue that quantum corrections
cannot change the form of eq.~\pref{LOW}, to all orders in
perturbation theory. What makes this argument tricky is the
observation that whatever form it takes, it {\it cannot} be based
on arguing $W$ is independent of $\tau$, due to the
$\tau$-dependence which already appears through the variables
$\cG^r$. This $\tau$ dependence arises because of the relative
factor of $\tau$ which distinguishes the NS-NS and RR fields,
$H^1_{(3)}$ and $H^2_{(3)}$, inside $G_{(3)}$. In its turn, the
$\tau$-dependence of $G_{(3)}$ can be traced to the statement that
the string coupling constant, $e^\phi$, is {\it not} the
loop-counting parameter for the low-energy 10D supergravity
lagrangian governing fluctuations about Type IIB vacua, even in
the string frame.\footnote{We thank J. Polchinski for emphasizing
this point.} This may be seen from the different factors of
$e^\phi$ which arise in the lagrangian of eq.~\pref{BoseSugra}, as
opposed to the corresponding action for heterotic vacua where the
dilaton appears as an overall factor of $e^{-2\phi}$ in the string
frame.

It is therefore convenient to organize the perturbative series
slightly differently, by re-scaling the string-frame fields as
follows:
\beq
e^\phi
\to \lambda \,  e^\phi, \qquad C_{(p)} \to \lambda^{-1} \,
C_{(p)}
\eeq
and $y^M \to \lambda^{-1/(p+1)} \, y^M$ (for the brane position,
$y^M$), for constant $\lambda$. Under this re-scaling we have
\beq
    F_{(p)} \to \lambda^{-1} \, F_{(p)} \qquad G_{(3)} \to
    \lambda^{-1} \, G_{(3)}\qquad   \sqrt{-\hat{g}} \, d^{(p+1)}y \to
    \lambda^{-1} \, \sqrt{-\hat{g}} \, d^{(p+1)}y
\eeq
and so the action of eq.~\pref{BoseSugra} satisfies $S \to
S/\lambda^2$. After performing this re-scaling we formally expand
all observables (and the low-energy 4D effective action) as a
series in $\lambda$, setting $\lambda = 1$ at the end of the
calculation. Since the action carries an overall factor of
$\lambda^{-2}$ the expansion in powers of $\lambda$ is simply the
loop expansion for the action of eq.~\pref{BoseSugra}.

Now comes the main point. Since $\lambda \to 1$ at the end of the
calculation, we do not claim that it is a good approximation to
work to any fixed order in powers of $\lambda$. In this regard
this makes the series in $\lambda$ unlike the {\it string} loop
expansion, for which successive terms are suppressed by powers of
the small quantity $e^\phi$. The $\lambda$ series simply
represents a reorganization of the string-loop expansion in powers
of $e^\phi$, in which terms are grouped according to their power
of $\lambda$ rather than their power of $e^\phi$. However, if we
can establish that $W$ is not corrected to {\it all} orders in
$\lambda$, then it also follows that it is not corrected to all
orders in $e^\phi$.

And it is simple to see that $W$ is not corrected from the
lowest-order result, eq.~\pref{LOW}, to any order in $\lambda$.
This is because under the above rescaling $\cG^r \to \lambda^{-1}
\cG^r$ and so the ratio $\cG^0/\cG^r$ is $\lambda$-independent. It
therefore follows from eq.~\pref{WSymForm} that the $R$ and PQ
symmetries imply that $w$ is $\lambda$ independent, and so $W$
transforms in precisely the same way as does $W_{\gvw}$: $W \to
\lambda^{-1} \, W$. It follows that $W$ is completely given by its
lowest-order approximation, $W = W_{\gvw}$, at lowest order in
$\alpha'$ but to all orders in the string coupling.

It remains to extend this result to all orders in $\alpha'$, and
this part of the argument follows much as for the heterotic
string. In particular we know that $W$ cannot depend on any
K\"ahler moduli to any order in $\alpha'$, because this is
precluded by a combination of holomorphy and shift symmetries for
the form fields which appear in these moduli. However the
dimensionless parameters which control the $\alpha'$ expansion are
powers of $\alpha'$ divided by the volume of various cycles of the
background Calabi-Yau space, and these volumes are all counted
among the Calabi-Yau's K\"ahler moduli. Since $W$ is independent
of these K\"ahler moduli, it follows that it is also uncorrected
to all orders in $\alpha'$. This establishes the desired result to
all orders in both the $\alpha'$ and string-coupling expansions.

At this point the hairs on the back of the reader's neck may be
bristling due to a visceral discomfort with re-ordering the loop
expansion, which is itself generically divergent.\footnote{We
thank Liam McAllister for helping us to drive a silver spike
through the heart of this particular demon.} We therefore make a
brief parenthetic aside at this point to help put the reader's
mind at rest. For these purposes recall that the loop expansion
(in either $\lambda$ or $e^\phi$) is asymptotic, and so the
vanishing of a quantity (like $W - W_{\gvw}$) to all orders in the
coupling is equivalent to the statement that the quantity vanishes
faster than any power of the relevant coupling as the coupling
goes to zero. Our goal is to sketch how this may be done for the
$e^\phi$ expansion given that it is true for the $\lambda$
expansion.

The main point is that although the classical action,
eq.~\pref{BoseSugra}, involves more than one overall power of
$e^\phi$, its dependence on $e^\phi$ is not arbitrarily
complicated. This is because the entire action corresponds either
to tree- or one-loop level in the string expansion (for the NS-NS
and RR terms respectively). Consequently, all of the contributions
which arise at any finite order in the $e^\phi$ expansion are
contained within a higher, but finite, number of terms in the
$\lambda$ series. If it is known that the contributions to any
particular quantity vanish to all orders of the $\lambda$
expansion, it is therefore possible to set up an inductive
argument which proves that the same is true to all orders in the
expansion in $e^\phi$.

\section{Discussion}

We provide in this note a derivation of the non-renormalization
theorem for those Type IIB vacua described by the
Gukov-Vafa-Witten superpotential, which is similar in spirit to
the well-known results for heterotic vacua in that it relies
purely on simple symmetry arguments (rather than on more detailed
properties of the calculus of perturbation theory in super-space).
We regard this to fill in an important missing step in the recent
arguments for the existence of discrete de Sitter type vacua for
Type IIB string theory.

We have made crucial use not only of the Peccei-Quinn symmetry
used in the original Dine-Seiberg proof for the heterotic case,
but also of the global $R$-symmetry as used  in Seiberg's proof
for the nonrenormalisation of $W$ in $N = 1$ supersymmetric field
theories \cite{seiberg, NRTheoremsSym}.

Notice however that even though in 4D these global symmetries
survive at all orders in perturbation theory, therefore
guaranteeing the valididty of the theorem to all orders, the
argument is a bit more complicated in string theory. The
complication arises because global symmetries of the effective,
leading-order, 10D action are known to be broken by combinations
of $\alpha'$ and string loop corrections and by the
compactification process. For instance, such $R$-symmetry breaking
terms are induced within topological string theory, including
terms like $\int F^{2g-2} R^2 d^4x$ where $F$ is the self-dual
part of the field strength of a graviphoton field in $N =2$
supergravity and $g$ is the genus of the worldsheet \cite{narain}.
Similar terms also arise for $N=1$ vacua.\footnote{We thank Nathan
Berkovits for calling these articles to our attention.}

These symmetry-breaking terms can be seen to be $F$-terms which
appear as higher derivative corrections to the effective action
and are therefore not captured by a superpotential. For the
validity of our proof, the only requirement that must be satisfied
is for these global symmetries to hold to leading order in the
$\alpha'$ expansion. The reason being that if the superpotential
is not renormalized at leading order in $\alpha'$ then the fact
that $W$ is independent of the K\"ahler moduli (which controls the
$\alpha'$ expansion) guarantees that $W$ will not be renormalized
in perturbation theory.

The main restriction of our analysis is the assumption of trivial
dilaton configurations, with $P_m \propto \partial_m \tau = 0$.
This assumption prevents us from directly extending our
conclusions to the generalizations of $W_\gvw$ which govern the
low-energy limit of $F$-theory compactifications
\cite{GVWFTheory}. Our argument does not directly apply in this
case because the 4D spurions, $\cP^r$, describing nonzero $P_m$
carry two units of $R$-charge, and so allow $W$ to depend on the
$R$-invariant ratios $\cP^r/\cG^0$. But this ratio scales like
$\lambda^2$ and so at face value does not forbid $W$ from
acquiring nontrivial changes at finite loop orders. 

Although we focus here on the superpotential, we notice in passing
that we expect similar arguments to apply to the holomorphic gauge
coupling function, $f_{ab}$. The simplest case to consider is that
for which the gauge fields live on a D7 brane, in which case the
lowest-order calculation gives a result independent of $\tau$:
\beq \label{LowestOrderf}
    f_{ab} = \sigma \, \delta_{ab} \,,
\eeq
where $\sigma$ is the appropriate volume modulus for the cycle on
which the D7 brane wraps. Since the PQ transformation does not
have an anomaly involving these gauge fields we know that $f_{ab}$
must be invariant under PQ transformations. Similarly,
$R$-invariance of the action requires that $f_{ab}$ should also be
invariant under the $R$ symmetry, and this implies
\beq
    f_{ab}(\varphi^i,\tau; \cG^r) = {\cal F}_{ab} \,
    \left(\varphi^i;\frac{\cG^r}{\cG^0} \right) \,,
\eeq
where ${\cal F}_{ab}$ is the most general $R$-invariant
combination built from $\varphi^i$ and $\cG^r$. {}From here on the
argument proceeds as before, by re-scaling fields by powers of
$\lambda$ so that $S \to S/\lambda^2$ and performing an expansion
in powers of $\lambda$. It follows that the function ${\cal
F}_{ab}$ receives no corrections in the string loop expansion
compared to the lowest-order result. Notice that this does {\it
not} preclude ${\cal F}_{ab}$ from differing from the lowest-order
result, eq.~\pref{LowestOrderf}, so long as the difference does
not depend on $\tau$. This agrees with the known direct
calculations \cite{fCorrections}, which find that the gauge
kinetic function receives nontrivial $\tau$-independent
corrections. It would be interesting to consider the case of gauge
couplings on D3 branes.

The main part of our argument involves the re-organization of the
string loop expansion in terms of the loop expansion of the
leading low-energy field theory. It is easier to show that no
corrections arise in this second expansion, and we argue that this
implies also the absence of corrections in the string loops. It is
clear that this kind of argument generalizes to other string vacua
besides the Type IIB case, for which both NS and Ramond states
also arise in the bosonic part of the low-energy field theory. In
particular, it would be of great interest to see whether similar
considerations are applicable to corrections to the lowest-order
F-theory superpotential, $W = \int G_4\wedge \Omega$, as well as
flux superpotentials for heterotic and type IIA strings.

\section*{Acknowledgements}
We thank useful conversations with N. Berkovits, R. Brustein, J.
Conlon, D. Cremades, S. de Alwis, M. Green, S. Kachru, E.
Kiritsis, J. Maldacena, L. McAllister, J. Polchinski, E.
Silverstein, A. Sinha and K. Suruliz. C.B.'s research is supported
by a grant from N.S.E.R.C. (Canada), as well as funds from
McMaster University, the Perimeter Institute and the Killam
Foundation. F.Q. is partially funded by PPARC a Royal Society
Wolfson  award and European Union Marie-Curie 6th Framework
programme - MRTN-CT-2004-503369 Quest for Unification. C.E. is
partially funded by EPSRC.

\end{document}